\documentclass[%
reprint,
 amsmath,amssymb,
 aps,
]{revtex4-1}
\usepackage[dvipdfmx]{graphicx}
\usepackage{braket}
\usepackage{bm}
\usepackage{amsmath}
\begin{document}
\title{A variety of clustering in $^{18}$O}
\author{T. Baba$^1$ and M. Kimura$^{2,3}$}
\affiliation{$^1$Kitami Institute of Technology, 090-8507 Kitami, Japan\\
$^2$Department of Physics, Hokkaido University, 060-0810 Sapporo, Japan\\
$^3$Reaction Nuclear Data Centre, Faculty of Science, Hokkaido University, 060-0810 Sapporo, Japan}
\date{\today}

\begin{abstract}
We investigate excited states of $^{18}$O by using the antisymmetrized molecular dynamics.
It is found that five different types of cluster states exist which we call ${}^{14}{\rm C}+\alpha$, higher-nodal ${}^{14}{\rm C}+\alpha$, two molecular states, and $4\alpha$ linear-chain.
The calculated $\alpha$-decay widths are compared with the observed data.
The higher-nodal ${}^{14}{\rm C}+\alpha$ cluster states reasonably agree with resonances reported by the recent experiments.
We predict that the $\alpha$-particle emission is dominant for the ${}^{14}{\rm C}+\alpha$ cluster states while the molecular states prefer the $^{6}$He emission.
\end{abstract}

\maketitle
\section{introduction}
Clustering is a well known phenomenon in light $N=Z$ self-conjugate nuclei.
The clustering of these nuclei shows striking characters such as an inversion doublet \cite{hori68}, large $\alpha$-decay widths \cite{free07}, and strong isoscalar monopole or dipole transitions \cite{chib16,enyo19}.
It has been suggested that more rich clustering can appear in $N\neq Z$ nuclei such as $^{9-12}$Be \cite{seya81,oert96,itag00}, $^{13-20}$C \cite{itag01,gree02,bohl03,itag04,ashw04,maru10,suha10,free14,ebra14,zhao15,
frit16,baba16,yama17,li17,baba17,ebra17,baba18}, $^{18-20}$O \cite{cuns81,gai83,desc85,gai91,curt02,furu08,oert10,avil14,naka18,yang19},
and $^{22-24}$Ne \cite{curt02,powe64,oliv93,yild06,kimu07,leva13,mare18} due to the degree-of-freedom of valence neutrons.
Although a number of works have provided information of the cluster structure in these nuclei,
it is not still sufficient to understand the nature of clustering in neutron-rich nuclei.

\begin{figure*}[b]
 \centering
 \includegraphics[width=1.0\hsize]{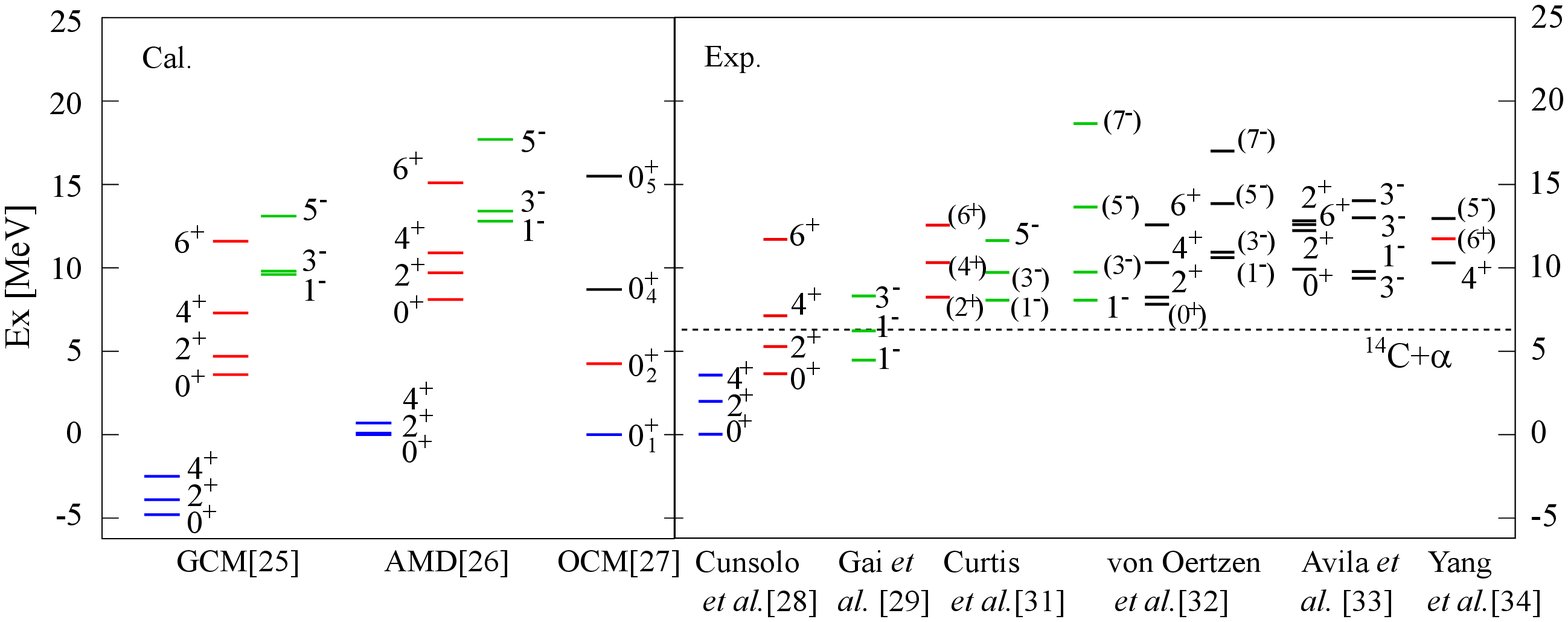}
 \caption{(color online) Theoretical energy spectra (left) and experimental energy spectra (right) of $^{18}$O. Blue, red, and green lines show the ground, positive-parity ${}^{14}{\rm C}+\alpha$ cluster, and negative-parity ${}^{14}{\rm C}+\alpha$ cluster bands, respectively.} 
 \label{fig:spc_pre}
\end{figure*}

$^{18}$O is a good example of the clustering in $N\neq Z$ nuclei because the core nucleus $^{16}$O has well-known ${}^{12}{\rm C}+\alpha$ cluster states \cite{hori68,libe80,enyo19}.
Figure \ref{fig:spc_pre} shows energy spectra of $^{18}$O reported by experimental and theoretical studies.
Cunsolo {\it et al.} suggested a positive-parity rotational band which consists of $0^+_2$; 3.63 MeV, $2^+_3$; 5.26 MeV, $4^+_2$; 7.11 MeV, $6^+_1$; 11.69 MeV \cite{cuns81}.
These states were tagged by $\alpha$-spectroscopic strengths.
Ever since, this band has long been established as the positive-parity ${}^{14}{\rm C}+\alpha$ band.
Gai {\it et al.} measured the electric dipole transition strength $B(E1)$ and suggested the negative-parity band of ${}^{14}{\rm C}+\alpha$ cluster, $1^-_1$ (4.45 MeV), $1^-_2$ (6.20 MeV), and $3^-_3$ (8.29 MeV) \cite{gai83,gai91}, which constitute the inversion doublet together with the positive-parity band suggested by Cunsolo {\it et al.}.
Curtis {\it et al.} measured a different ${}^{14}{\rm C}+\alpha$ cluster bands, $1^-$; 8.04 MeV, $2^+$; 8.22 MeV, $3^-$; 9.70 MeV, $4^+$; 10.29 MeV, $5^-$; 11.62 MeV, by the ${}^{14}{\rm C}({}^{18}{\rm O}, {}^{14}{\rm C}\alpha){}^{14}{\rm C}$ reaction \cite{curt02}.
Subsequently, von Oertzen {\it et al.} interpreted these bands as follows \cite{oert10}.
The band ($0^+_2$, $2^+_3$, $4^+_2$, $6^+_1$) observed by Cunsolo {\it et al.} and Gai {\it et al.} is the ${}^{14}{\rm C}+\alpha$ cluster, while another band observed by Curtis {\it et al.} is a molecular ${}^{12}{\rm C}+\alpha+2n$ structure.
In addition, they newly predicted negative-parity bands for each positive-parity band, although the spin-parities of almost all states were tentative.
Based on ${}^{14}{\rm C}+\alpha$ elastic scattering, Avila {\it et al.} obtained the detailed spectroscopic information including partial $\alpha$- and $n$-decay widths \cite{avil14}.
They reported very large $\alpha$-decay widths for several resonances.
Moreover, they excluded the negative-parity states suggested by von Oertzen {\it et al.} because they observed small $\alpha$-decay widths for these states.
More recently, Yang {\it et al.} performed a multi-nucleon transfer experiment ${}^{9}{\rm Be}({}^{13}{\rm C}, \alpha+{}^{14}{\rm C})\alpha$ \cite{yang19}.
They observed new resonances and $\alpha$-branching ratios for the previously reported states.
In particular, they were able to determine the 11.7 MeV state as a $6^+$ state owing to the better resolution, which had been remained the unsolved spin-parity ($6^+$ or $5^-$).

Theoretically, Descouvemont and Baye investigated ${}^{14}{\rm C}+\alpha$ states using the generator coordinate method (GCM) \cite{desc85}.
They used asymmetric ${}^{14}{\rm C}+\alpha$($0^+_1$) and ${}^{14}{\rm C}+\alpha$($2^+_1$) wave functions and compared the results with experiments.
As a result, the $K=0^+$ band ($0^+_2$; 3.6 MeV, $2^+_3$; 4.7 MeV, $4^+_2$; 7.3 MeV, $6^+_1$; 11.6 MeV) and the $K=0^-$ band ($1^-_3$; 9.6 MeV, $3^-_3$; 9.8 MeV, $5^-_2$; 13.1 MeV) were predicted.
Furutachi {\it et al.} studied $^{18}$O using the antisymmetrized molecular dynamics (AMD) \cite{furu08}.
They showed that the ground band and molecular ${}^{14}{\rm C}+\alpha$ band roughly agreed with experimental ones but there are still energy gaps.
Quite recently, Nakao {\it et al.} analyzed Coulomb shifts of the ${}^{14}{\rm C}+\alpha$ cluster by applying the orthogonality condition model (OCM) \cite{naka18}.
This calculation nicely reproduced the observed $0^+$ state including higher-lying states.
In addition, the Thomas-Ehrman shift and the monopole transition are proposed as the probe of clustering.

Although the many experimental and theoretical studies have been performed, there are two problems to be solved.
First, the ${}^{12}{\rm C}+\alpha+2n$ molecular states suggested by von Oertzen {\it et al.} have not yet been investigated theoretically.
In the case of ${}^{12}{\rm Be}$ \cite{ito08}, it is shown that the $\alpha+\alpha+4n$ configuration appears at the lower excitation energy than various two-body cluster states such as ${}^{8}{\rm He}+\alpha$, ${}^{6}{\rm He}+{}^{6}{\rm He}$, and ${}^{7}{\rm He}+{}^{5}{\rm He}$.
The former is interpreted as the molecular orbital structure analog to the covalent bond where the four valence neutrons locates around the clusters simultaneously, whereas the latter is interpreted as the ionic configuration where the four valence neutrons are trapped in either of $\alpha$-cores.
If the ${}^{14}{\rm C}+\alpha$ state is interpreted as the "ionic" state, the molecular orbital configuration ${}^{12}{\rm C}+\alpha+2n$ should be also expected.
Second, there is lack of quantitative calculation to compare with the observations.
As mentioned above, the observables such as B(E1) and $\alpha$-decay widths have been already measured.
Therefore, we need to calculate the B(E1) and $\alpha$-decay widths quantitatively to establish the clustering in ${}^{18}{\rm O}$.
In addition, the inversion doublet is essential to prove the asymmetric cluster structure, but the assignment of the negative-parity ${}^{14}{\rm C}+\alpha$ bands are controversial between the experiments.

In this paper, we investigate excited states of ${}^{18}{\rm O}$ by using the AMD.
For quantitative comparison of the excitation energy, we improve a effective nucleon-nucleon interaction and wave functions compared with the previous AMD calculation \cite{furu08}.
To search for the molecular states, we calculate higher excited states.
As the result, we suggest that the five cluster configurations exist.
Moreover, we compare the energy spectra and $\alpha$-decay widths with the experiments.
Calculated higher-nodal ${}^{14}{\rm C}+\alpha$ cluster states reasonably agree with resonances reported by the recent experiments.
Also, it is found that the ${}^{14}{\rm C}+\alpha$ cluster states decay by $\alpha$-particle emission while the ${}^{12}{\rm C}+\alpha+2n$ molecular states decay by $^{6}$He emission.

\section{Theoretical framework}
In this work, we use the microscopic $A$-body Hamiltonian written as 
\begin{align}
 \hat{H} &= \sum_{i=1}^A \hat{t}_i - \hat{t}_{c.m.} + \sum_{i<j}^A \hat{v}^N_{ij} + \sum_{i<j}^Z \hat{v}^C_{ij}, \label{eq:h}
\end{align}
where $\hat{t}_i$, $\hat{t}_{c.m.}$, $\hat{v}^N$, and $\hat{v}^C$ are the kinetic energy per nucleon, kinetic energy of the center-of-mass, nucleon-nucleon interaction and Coulomb interaction, respectively.
The Gogny D1S \cite{gogn91} is used as the effective nucleon-nucleon interaction .

The AMD wave function $\Phi_{\rm AMD}$ is represented by a Slater determinant of single
particle wave packets, 
\begin{align}
 \Phi_{\rm AMD} ={\mathcal A} \{\varphi_1,\varphi_2,...,\varphi_A \}
 =\frac{1}{\sqrt{A!}}\mathrm{det}[\varphi_{i}({\bm r}_j)].
  \label{eq:amdwf}  
\end{align}
Here, $\varphi_i$ is the single particle wave packet which is a direct product of the deformed
Gaussian spatial part \cite{kimu04}, spin ($\chi_i$) and isospin ($\xi_i$) parts,  
\begin{align}
 \varphi_i({\bm r}) &= \phi_i({\bm r})\otimes \chi_i\otimes \xi_i, \label{eq:singlewf}\\
 \phi_i({\bm r}) &= \prod_{\sigma=x,y,z}\Bigl(\frac{2\nu_\sigma}{\pi}\Bigr)^{1/4}\exp\biggl\{-\nu_\sigma\Bigl(r_\sigma -\frac{Z_{i\sigma}}{\sqrt{\nu_\sigma}}\Bigr)^2\biggr\}, \\
 \chi_i &= a_i\chi_\uparrow + b_i\chi_\downarrow,\quad
 \xi_i = {\rm proton} \quad {\rm or} \quad {\rm neutron}.\nonumber
\end{align}
The centroids of the Gaussian wave packets $\bm Z_i$, the direction of nucleon spin $a_i, b_i$,
and the width parameter of the deformed Gaussian $\nu_\sigma$ are variables determined by the variational calculation.
In this calculation, we perform the variational calculation with a constraint potential on the quadrupole deformation parameter $\beta$,
\begin{align}
 E'^\pi = \frac{\braket{\Phi^\pi|{H}|\Phi^\pi}}{\braket{\Phi^\pi|\Phi^\pi}} 
 + v_\beta(\braket{\beta} - \beta_0)^2.
\end{align}
After the variational calculation, the eigenstate of the total angular momentum is projected out.

We perform the generator coordinate method by employing the quadrupole deformation parameter $\beta$ as the generator coordinate.
The wave functions $\Phi^{J\pi}_{MKi}$ are superposed, 
\begin{align}
 \Psi^{J\pi}_{M\alpha} = \sum_{Ki}g^{J}_{Ki\alpha}\Phi^{J\pi}_{MKi},\label{eq:gcmwf}
\end{align}
where the coefficients $g^{J}_{Ki\alpha}$ and eigenenergies $E^{J\pi}_\alpha$ are obtained by solving the Hill-Wheeler equation \cite{hill54}.
To discuss the dominant configuration in $\Psi_{Mn}^{J^\Pi}$, we calculate the overlap between $\Psi_{Mn}^{J^\Pi}$ and the basis wave function $\Phi^{J^\Pi}_{MK}(\beta_i)$, 
\begin{align}
 |\braket{\Phi^{J^\Pi}_{MK}(\beta)|\Psi^{J^\Pi}_{Mn}}|^2/
 \braket{\Phi^{J^\Pi}_{MK}(\beta)|\Phi^{J^\Pi}_{MK}(\beta)}. \label{eq:gcmovlp}
\end{align}

Using the GCM wave functions, we estimate the $\alpha$-decay width from the reduced width amplitude (RWA).
In order to calculate the RWA, we employ the Laplace expansion method given in Ref. \cite{chib17}.
The reduced width $\gamma_{l}$ is given by the square of the RWA,
\begin{align}
 \gamma^2_{l}(a) = \frac{\hbar^2}{2\mu a}|ay_{l}(a)|^2,
\end{align}
and the partial $\alpha$-decay width is a product of the reduced width and the penetration factor $P_l(a)$,
\begin{align}
 \Gamma_{l} &= 2P_l(a)\gamma^2_{l}(a), \quad
 P_l(a) = \frac{ka}{F^2_l(ka)+G^2_l(ka)}, 
\end{align}
where $P_l$ is given by the Coulomb regular and irregular wave functions $F_l$ and $G_l$.
Here, the channel radius $a$ is chosen as $5.2$ fm which is same with those used in Refs. \cite{avil14,yang19}.
The wave number $k$ is determined by the decay $Q$-value and the reduced mass $\mu$ as $k=\sqrt{2\mu E_Q}$.
A dimensionless $\alpha$-reduced width is defined by the ratio of the reduced width to Wigner limit $\gamma^2_{W}$,
\begin{align}
 \theta^2_{l}(a) = \frac{\gamma^2_{l}(a)}{\gamma^2_{W}(a)}, \quad
 \gamma^2_{W}(a) = \frac{3\hbar^2}{2\mu a^2},
\end{align}
and the spectroscopic factor $S$ is defined by the integral of the RWA,
\begin{align}
 S = \int^{\infty}_0 r^2|y_{l}(r)|^2dr. \label{eq:sfactor}
\end{align}

To investigate the valence-neutron properties, we calculate single-particle orbits of the intrinsic wave function.
The calculation is referred in our previous work \cite{baba16}.
Using the single-particle orbit $\widetilde{\phi}_s$, we discuss the amount of the positive-parity component, 
\begin{align}
 p^+ = |\langle \widetilde{\phi}_s|\frac{1+P_x}{2}| \widetilde{\phi}_s\rangle|^2, \label{eq:sp1}
\end{align}
and angular momenta in the intrinsic frame,
\begin{align}
 j(j+1)&= \langle \widetilde{\phi}_s|\hat{j}^2| \widetilde{\phi}_s\rangle, \quad
 |j_z| = \sqrt{\langle \widetilde{\phi}_s|\hat{j}_z^2| \widetilde{\phi}_s\rangle},\label{eq:sp2}\\
 l(l+1)&= \langle \widetilde{\phi}_s|\hat{l}^2| \widetilde{\phi}_s\rangle, \quad
 |l_z| = \sqrt{\langle \widetilde{\phi}_s|\hat{l}_z^2| \widetilde{\phi}_s\rangle}.\label{eq:sp3}
\end{align}

\section{Results}
\subsection{Energy surface and density distribution}
\begin{figure}[h]
 \centering
 \includegraphics[width=1.0\hsize]{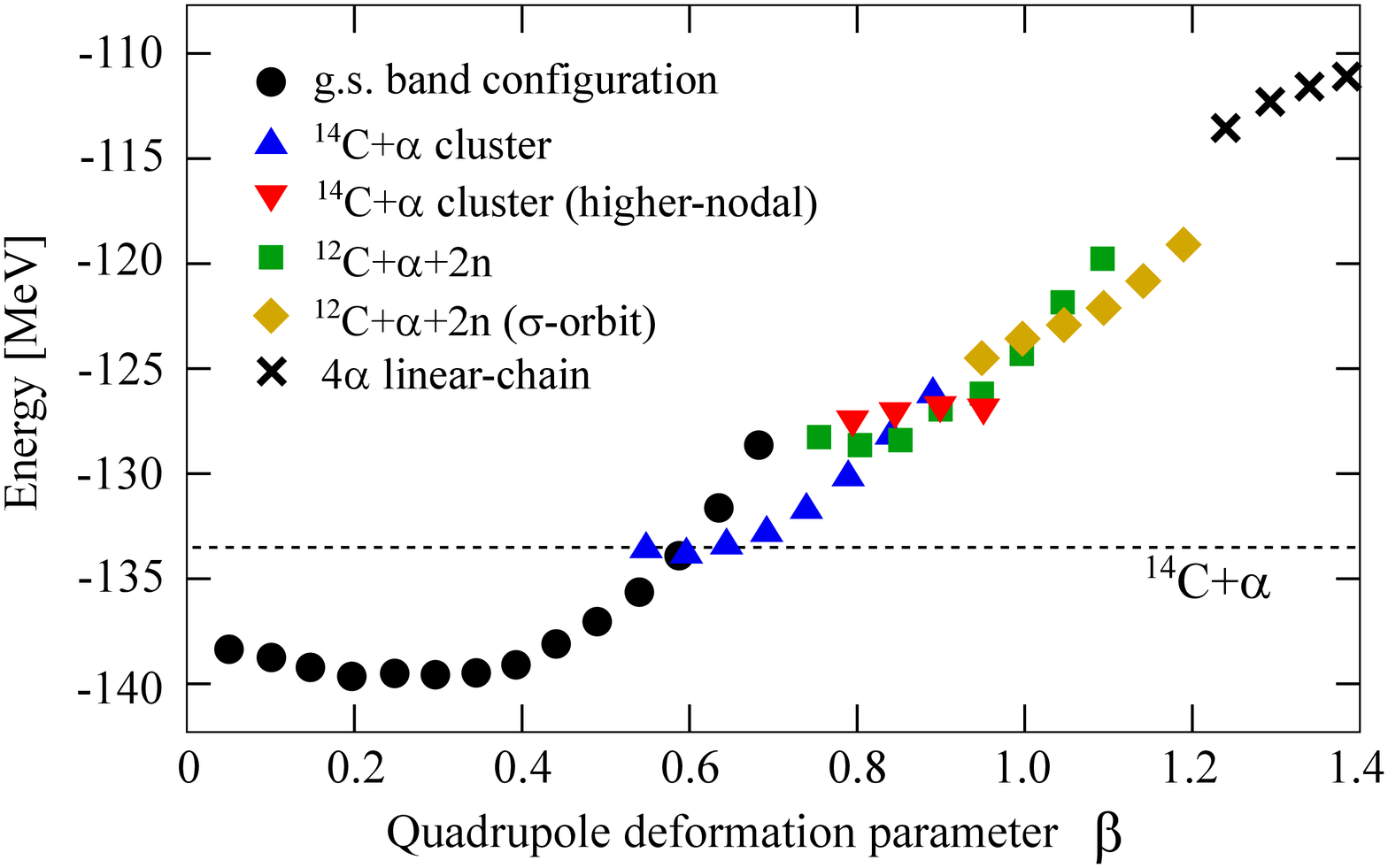}
 \caption{(color online) Energy curves as a function of a quadrupole deformation parameter $\beta$ of $J^\pi = 0^+$ in $^{18}$O. It is appeared that six configurations shown by density distributions in Fig. \ref{fig:dens+}.} 
 \label{fig:surf+}
\end{figure}
\begin{figure}[h]
 \centering
 \includegraphics[width=1.0\hsize]{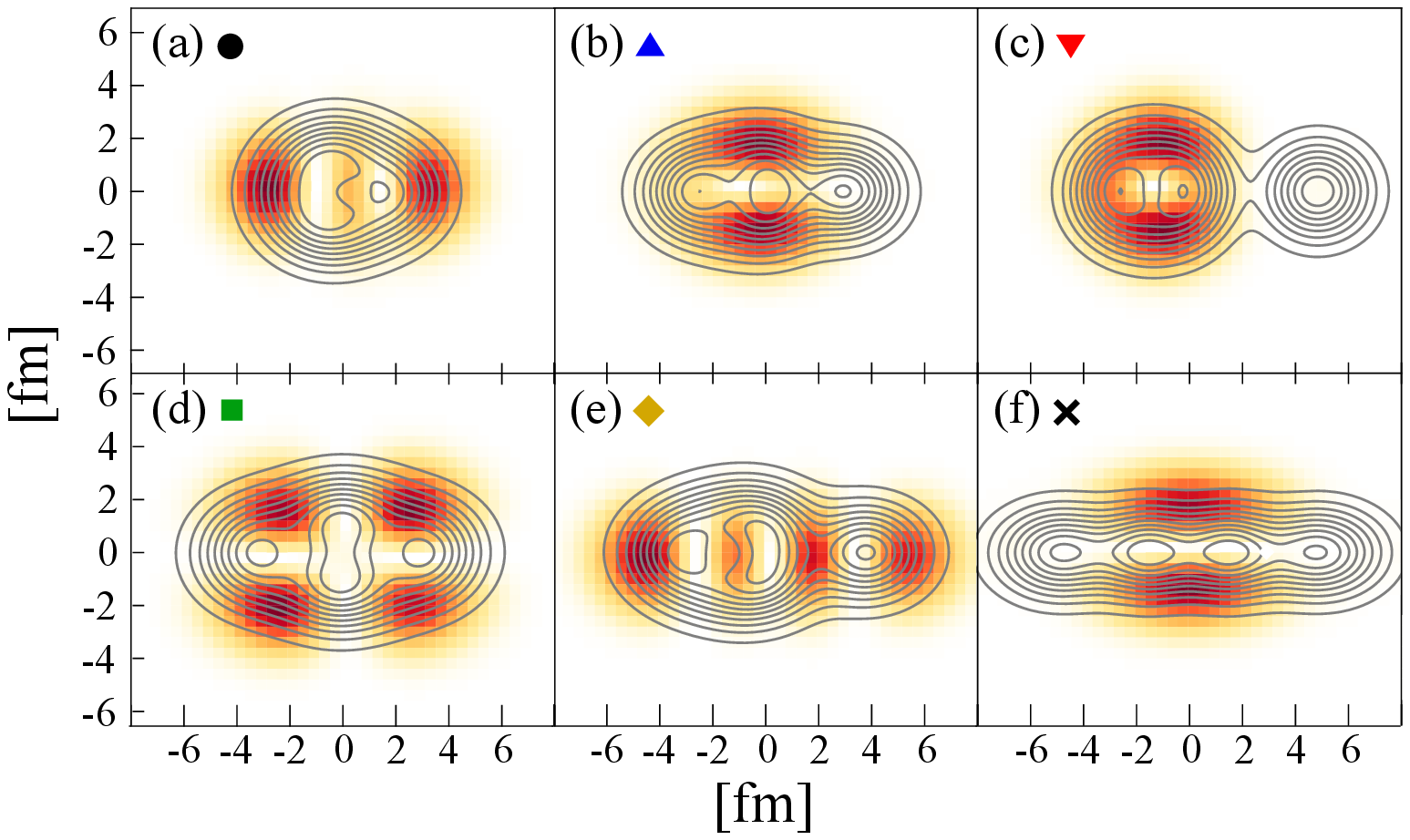}
 \caption{(color online) Density distributions of positive-parity states in $^{18}$O obtained by the variational calculation. Contour lines show the proton density distributions and color plots show the single particle orbits occupied by two valence neutrons. See the text for the detail of each panel.} 
 \label{fig:dens+}
\end{figure}

Figure \ref{fig:surf+} shows energy curves for $J^\pi = 0^+$ states obtained by the $\beta$-constraint variational calculation.
On these curves, six different structures appear whose density distributions are illustrated in Fig. \ref{fig:dens+}.
There are also other local energy minima with different structures above these energy curves. However, they do not have prominent cluster structure, and hence, are not shown in this figure.
We focus on and discuss these six structures by referring their density distributions and properties of valence neutron orbits listed in Table. \ref{tab:spo+}.

The lowest energy configuration shown by circles has the minimum at $E=-139.6$ MeV and $\beta=0.2$.
As seen in Fig. \ref{fig:dens+} (a), this configuration has no pronounced clustering and it becomes the most dominant component of the ground band.
The properties of the valence neuron in Table. \ref{tab:spo+} (a) shows two valence neutrons approximately occupy $(d_{5/2})^2$ orbit, that is, $j\approx 5/2$ and $l\approx 2$.

In $\beta=0.6\sim0.8$ region, another configuration shown by blue triangles becomes the local energy minimum.
Figure \ref{fig:dens+} (b) shows that it has the spatially separated $\alpha$ and $^{14}$C clusters.
In our calculation, positive-parity ${}^{14}{\rm C}+\alpha$ cluster configuration is little different from that of previous AMD work in Ref. \cite{furu08}.
Valence neutrons occupy the $p$-orbit perpendicular to the symmetry axis, while they occupy the orbit parallel to the symmetry axis in previous work.

In $\beta=0.8\sim1.0$ region, two different configurations shown by red triangles and green squares are almost degenerated.
The red triangles show the well developed $\alpha$ and $^{14}$C clustering illustrated by Fig. \ref{fig:dens+} (c).
On the other hand, the configuration shown by the green squares has a symmetric configuration illustrated by Fig. \ref{fig:dens+} (d).
It seems that two valence neutrons distribute all over the nucleus.
We consider that this configuration corresponds to ${}^{12}{\rm C}+\alpha+2n$ molecular states suggested by von Oertzen {\it et al} \cite{oert10} although our calculation does not show the clear ${}^{12}{\rm C}$ and $\alpha$ core.
The properties of the valence neuron in Table. \ref{tab:spo+} (d) show that the valence neutrons occupy the $d$-orbit, which behaves like the molecular orbit.

In $\beta=1.0\sim1.2$ region, another molecular configuration denoted by yellow diamonds becomes the yrast states.
Figure \ref{fig:dens+} (e) displays well developed $\alpha$ clustering but a different orbit of the valence neutrons which locates parallel to the symmetry axis.
From the single-particle properties in Table. \ref{tab:spo+} (e), it is found that the valence neutrons occupy the $\sigma$-orbit, that is, $|j_z|\approx1/2$ and $|l_z|\approx0$.
This configuration is quite similar to the $\sigma$-orbit predicted for $^{22}$Ne \cite{kimu07}.

In extremely deformed region $\beta>1.2$, the exotic clustering is realized, which is denoted by cross symbols.
Figure \ref{fig:dens+} (f) shows the linear alignment of 4$\alpha$ particles.
In addition, two valence neutrons occupy the $\pi$-orbit ($|j_z|\approx3/2$ and $|l_z|\approx1$) from Table. \ref{tab:spo+} (f).
Thus, we show the $\pi$-bond linear-chain configuration exists in 4$\alpha$ system, similar to $^{14}$C \cite{suha10,baba16,baba17}.
\begin{table}[h]
\caption{Properties of valence neutron orbits shown in Fig. \ref{fig:dens+}. Each column show the single particle energy $\varepsilon$ in MeV, the amount of the positive-parity component $p^+$ and the angular momenta (see Eqs. (\ref{eq:sp1})-(\ref{eq:sp3})).}
\label{tab:spo+}
\begin{center}
 \begin{ruledtabular}
  \begin{tabular}{ccccccc} 
	orbit & $\varepsilon $ & $p^+$ & $j$ & $|j_{z}|$ & $l$ & $|l_{z}|$ \\ \hline
	(a) & $-6.75$ & 1.00 & 2.31 & 0.50 & 1.86 & 0.44 \\
	(b) & $-9.55$ & 0.07 & 0.96 & 0.50 & 1.20 & 0.97 \\
	(c) & $-8.78$ & 0.29 & 1.07 & 0.50 & 1.38 & 0.90 \\
	(d) & $-5.72$ & 0.85 & 2.62 & 1.30 & 2.38 & 1.01 \\
	(e) & $-5.22$ & 0.05 & 2.61 & 0.54 & 2.38 & 0.35 \\
	(f) & $-7.26$ & 0.07 & 1.97 & 1.49 & 1.58 & 0.99 \\ 
  \end{tabular}
  \end{ruledtabular}
\end{center}
\end{table}

\begin{figure}[h]
 \centering
 \includegraphics[width=1.0\hsize]{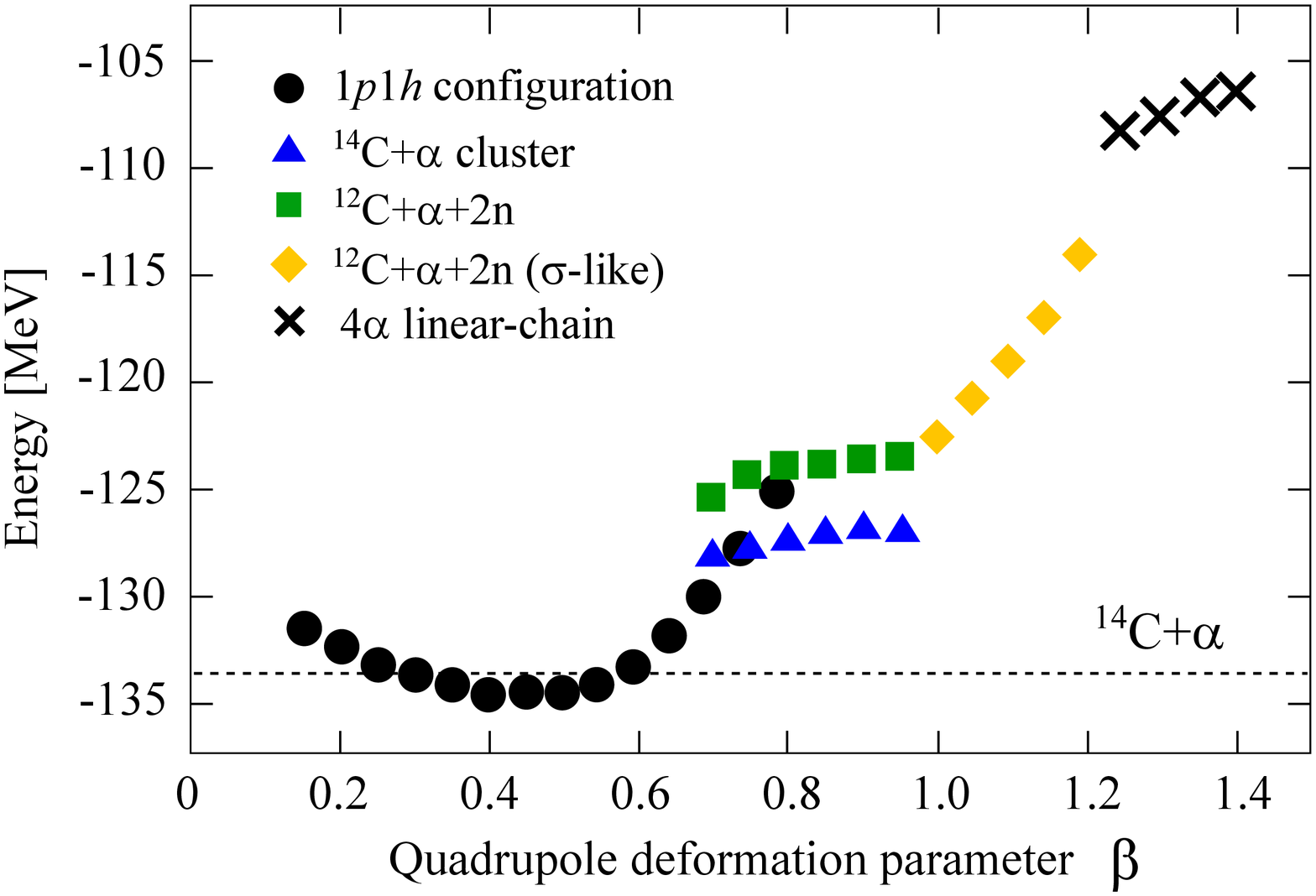}
 \caption{(color online) Energy curves as a function of a quadrupole deformation parameter $\beta$ of $J^\pi = 1^-$ in $^{18}$O. It is appeared that five configurations shown by density distributions in Fig. \ref{fig:dens-}.} 
 \label{fig:surf-}
\end{figure}
\begin{figure*}[h]
 \centering
 \includegraphics[width=1.0\hsize]{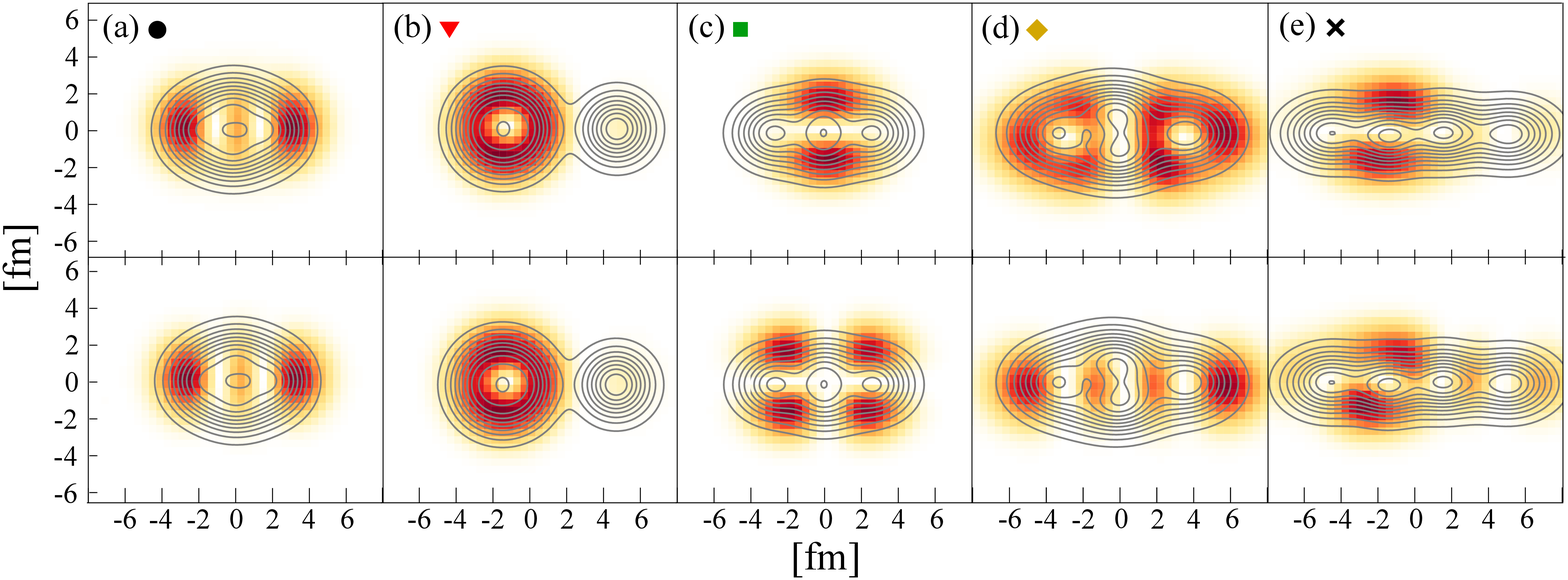}
 \caption{(color online) Density distributions of negative-parity states in $^{18}$O obtained by the variational calculation. The lower panels show the most weakly bound neutron, while the upper panel  show the other valence neutron. See the text for the detail of each panel.} 
 \label{fig:dens-}
\end{figure*}

Figure \ref{fig:surf-} shows energy curves for $J^\pi = 1^-$ states in $^{18}$O.
On this figure, five different structures appear, which are illustrated by density distributions in Fig. \ref{fig:dens-} and properties of valence neutron orbits listed in Table. \ref{tab:spo-}.

The energy minimum of the $1^-$ state with negative-parity is located at $\beta=0.40$ with the binding energy $-134.57$ MeV shown by circles, which has the density distribution described in Fig.\ref{fig:dens-} (a).
From Table. \ref{tab:spo-}, it is found that valence neutrons occupy the $(d_{5/2})^2$ orbit which is same as the ground state.
In this configuration, the most weakly bound proton is excited into the $d_{5/2}$ orbit ({\it i.e.} $1p1h$ configuration $\pi (p_{1/2})^{-1}(d_{5/2})^1$), so that the negative-parity is attained.

In $\beta=0.8\sim1.0$ region, two different configurations shown by blue triangles and green squares appear, which is similar to the positive-parity.
The blue triangles show the pronounced ${}^{14}{\rm C}+\alpha$ clustering illustrated by Fig. \ref{fig:dens-} (b).
This configuration is almost same as that of positive-parity.
Therefore, it is the counterpart of the inversion doublet.
On the other hand, the configuration shown by the green squares is the molecular configuration but little different from that of the positive-parity.
The most weakly bound neutron shown in the lower panel of Fig. \ref{fig:dens-} (c) is same as that of the positive-parity molecular state, namely, the $d$-orbit.
The other valence neutron, however, has different properties $p^+=0.01$, $|j_z|=0.54$ and $|l_z|=1.00$ in Table. \ref{tab:spo-} (c).
These properties correspond to the $\pi$-orbit.

In $\beta=1.0\sim1.2$ region, the steep curve shown by yellow diamonds becomes the lowest energy configuration.
Figure \ref{fig:dens-} (d) shows a similar configuration to the ${}^{12}{\rm C}+\alpha+2n$ molecular states with the $\sigma$-orbit.
Actually, the most weakly bound neutron occupy the $\sigma$-orbit because $|j_z|\approx1/2$ and $|l_z|\approx0$.
However, the other valence neutron does not show a clear molecular orbit because of the parity mixing ($p^+=0.51$).

Similar to the positive-parity, the 4$\alpha$ linear-chain configuration appears at $\beta>1.2$ with rather high excitation energy.
From Fig. \ref{fig:dens-} (e), it seems that the two valence neutrons occupy the $\pi$-orbit.
However, their properties show that the parity mixing occurs even in this configuration.
Note that these orbits locate around left 3$\alpha$ showing the ${}^{14}{\rm C}+\alpha$ correlation.
\begin{table}[h]
\caption{Properties of two valence neutron orbits shown in Fig. \ref{fig:dens-}.} 
\label{tab:spo-}
\begin{center}
 \begin{ruledtabular}
  \begin{tabular}{ccccccc} 
	orbit & $\varepsilon $ & $p^+$ & $j$ & $|j_{z}|$ & $l$ & $|l_{z}|$ \\ \hline
	(a) & $-8.77$ & 0.99 & 2.32 & 0.56 & 1.87 & 0.50 \\
	    & $-6.88$ & 0.99 & 2.29 & 0.63 & 1.86 & 0.49 \\ 
	(b) & $-9.11$ & 0.38 & 1.03 & 0.50 & 1.38 & 0.83 \\
	    & $-9.09$ & 0.38 & 1.03 & 0.50 & 1.38 & 0.83 \\ 
	(c) & $-9.39$ & 0.01 & 0.94 & 0.54 & 1.14 & 1.00 \\
	    & $-5.63$ & 0.99 & 2.63 & 1.50 & 2.24 & 1.04 \\ 
	(d) & $-5.19$ & 0.51 & 2.95 & 1.07 & 2.73 & 0.78 \\
	    & $-4.79$ & 0.02 & 3.02 & 0.57 & 2.84 & 0.26 \\ 
	(e) & $-6.59$ & 0.26 & 2.38 & 1.44 & 2.07 & 0.96 \\
	    & $-5.36$ & 0.56 & 2.74 & 1.30 & 2.48 & 0.86 \\ 
  \end{tabular}
  \end{ruledtabular}
\end{center}
\end{table}

\subsection{Energy spectrum}
\begin{figure}[h]
 \centering
 \includegraphics[width=1.0\hsize]{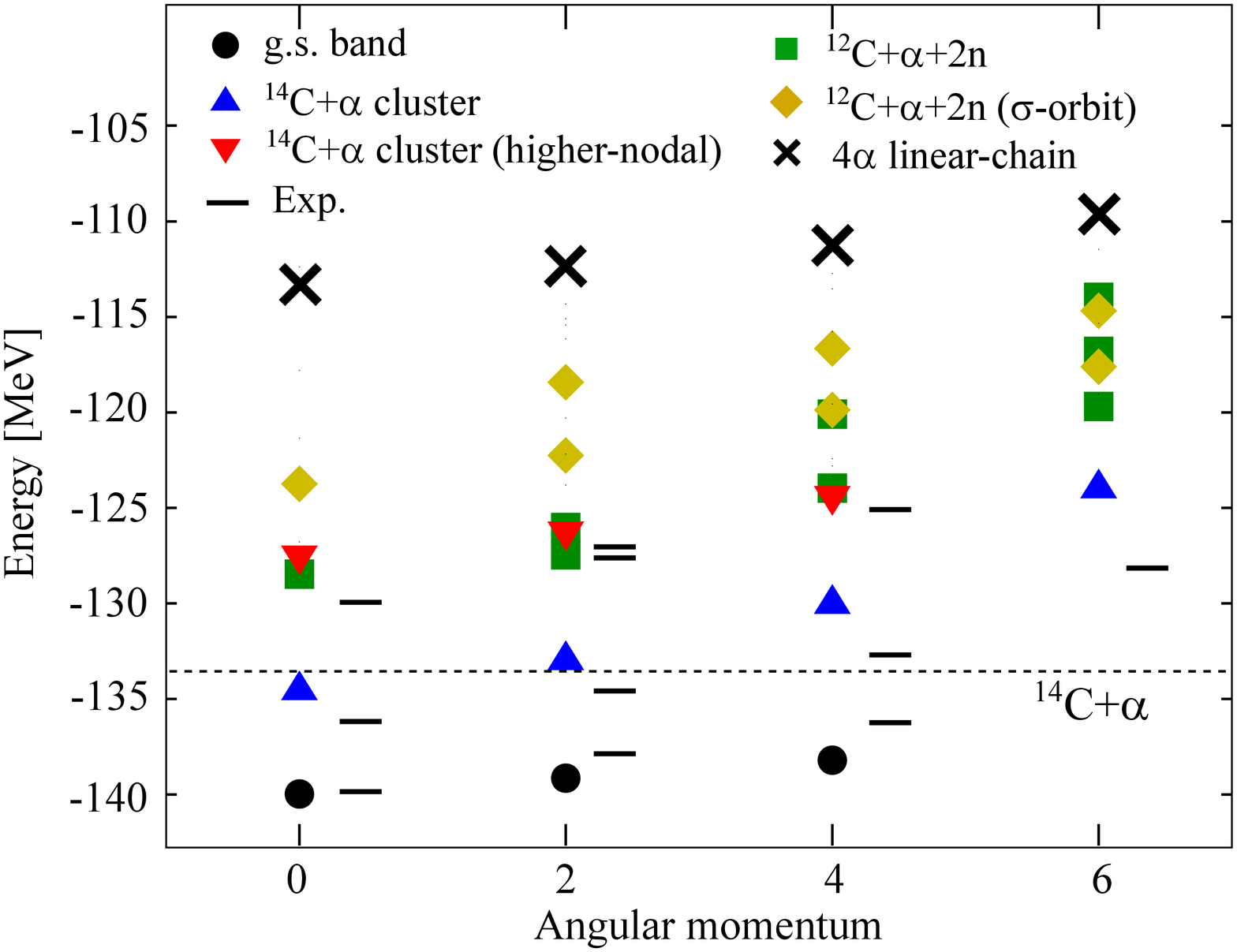}
 \caption{(color online) Positive-parity energy spectrum of $^{18}$O up to $J^\pi=6^+$. 
 Lines show the observed data \cite{cuns81,avil14,endt93} and any other symbols are notated in the same manner as Figs. \ref{fig:surf+} and \ref{fig:dens+}.
 The non-cluster states are not shown.} 
 \label{fig:spec+}
\end{figure}
Figure \ref{fig:spec+} shows the positive-parity spectrum up to $J^\pi=6^+$ state obtained by the GCM calculation.
We classified the obtained states to six bands and other non-cluster states based on the configurations discussed in the previous section.
This classification is based on their overlap with the basis wave functions defined by Eq. (\ref{eq:gcmovlp}).
Table \ref{tab:width+} lists the member states of these bands and compares with the observed data.

The member states of the ground band shown by circles in Fig. \ref{fig:spec+} are dominantly composed of the basis wave function shown in Fig. \ref{fig:dens+} (a).
In fact, the ground state has the largest overlap with this basis that amounts to 0.98.
The calculated binding energy is $-139.97$ MeV that nicely agrees with the observed value ($-139.81$ MeV).
Due to the improvement of the wave functions and effective interaction, the moment-of-inertia of the ground band is smaller than that of the previous AMD framework \cite{furu08}, as a result, the excitation energies of the $2^+_1$ and $4^+_1$ states are also reasonably improved.

\begin{figure}[h]
 \centering
 \includegraphics[width=1.0\hsize]{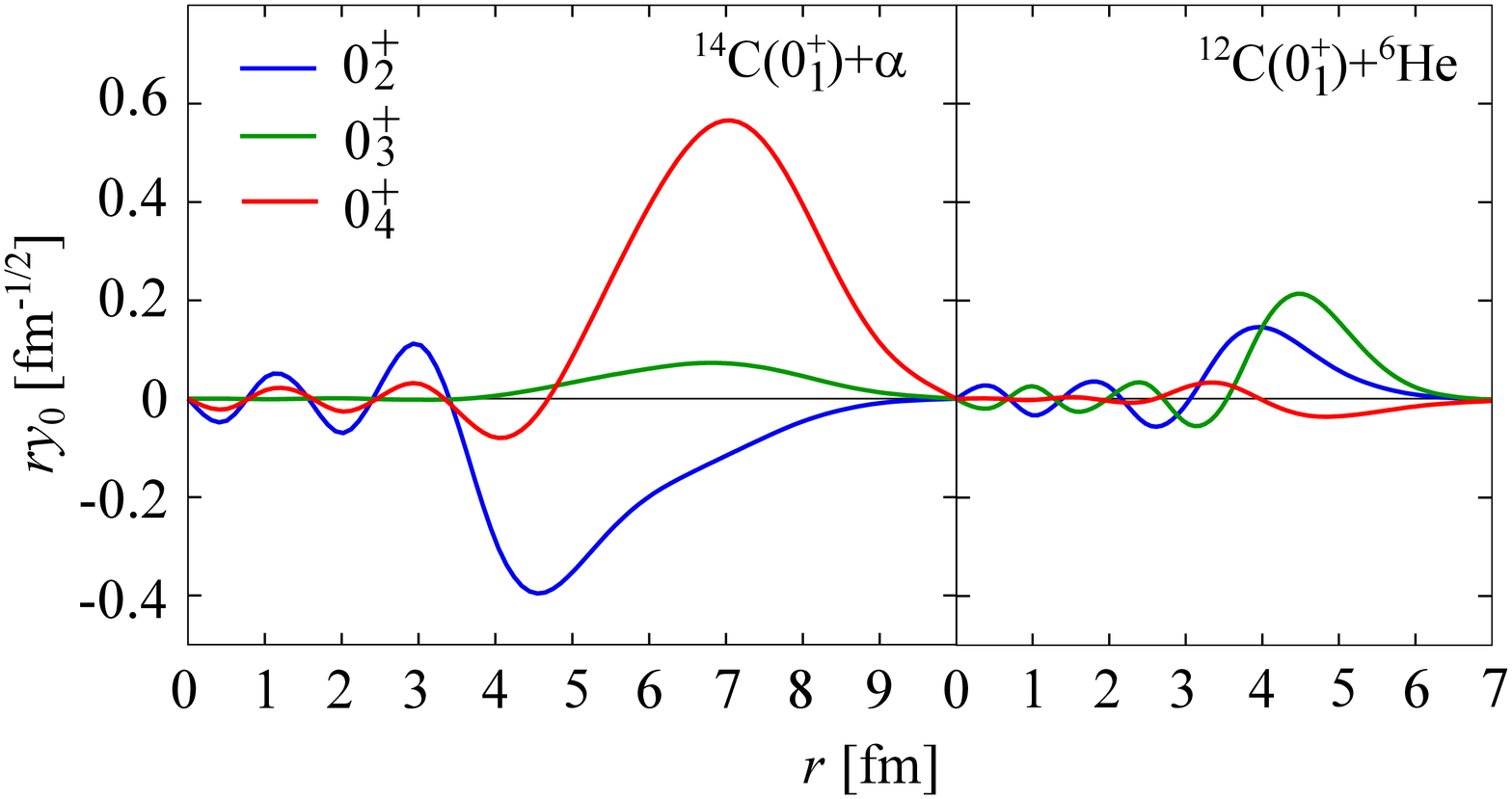}
 \caption{(color online) Reduced width amplitude as a function of distance $r$ for $0^+_2$ (blue), $0^+_3$ (green), and $0^+_4$ (red) states as the ${}^{14}{\rm C}+\alpha$ (left) and ${}^{12}$C+$^6$He (right) channels. It is assumed that the spins of $^{4,6}$He and $^{12,14}$C, and relative angular momentum between them are zero.} 
 \label{fig:rwa}
\end{figure}

The ${}^{14}{\rm C}+\alpha$ cluster configuration generates a rotational band denoted by blue triangles at 5.44 MeV near the ${}^{14}{\rm C}+\alpha$ threshold.
The bandhead state $0^+_2$ has the largest overlap with the basis wave function shown in Fig.\ref{fig:dens+} (b) which amounts to 0.94.
Excitation energies of this band are closer to those of the observation than the previous AMD framework although they are still overestimated.
The band shown by red triangles is composed of the basis wave function in Fig. \ref{fig:dens+} (c).
The bandhead state $0^+_4$ has the largest overlap with this basis which amounts to 0.71.
To make the difference between $0^+_2$, $0^+_3$, and $0^+_4$ states clear, their reduced width amplitudes are shown in Fig. \ref{fig:rwa}.
In the ${}^{14}{\rm C}+\alpha$ channel (left panel), the $0^+_2$ state has four nodes ($n=4$) while the $0^+_4$ state has five nodes ($n=5$).
Therefore, we conclude that the $0^+_4$ state is the higher-nodal ${}^{14}{\rm C}+\alpha$ state, which corresponds the $0^+_4$ state in OCM calculation \cite{naka18} and the $0^+$ state observed in Ref. \cite{avil14}.
In addition, the candidates for $2^+$ and $4^+$ states which have rather large $\alpha$-decay widths are also observed, and listed in Table. \ref{tab:width+}.

The ${}^{12}{\rm C}+\alpha+2n$ molecular configurations, denoted by green squares and yellow diamonds, generate two rotational bands respectively.
The former band generates the $K^\pi=0^+$ band built on $0^+_3$ state at 11.34 MeV, and the $K^\pi=2^+$ band built on $2^+_6$ state at 13.78 MeV.
The bandhead $0^+_3$ state has the largest overlap with the wave function shown in Fig.\ref{fig:dens+} (d) that amounts to 0.97.
The $\sigma$-orbit configuration generates the $K^\pi=0^+$ band built on $0^+_6$ state at 16.07 MeV, and the $K^\pi=2^+$ band built on $2^+_{11}$ state at 21.39 MeV.
The bandhead $0^+_6$ state has the largest overlap with the wave function shown in Fig.\ref{fig:dens+} (e) that amounts to 0.87.
We consider that the green squares band corresponds the band suggested by von Oertzen {\it et al} \cite{oert10} although our calculation overestimates their excitation energies. 
Note that the molecular states ${}^{12}{\rm C}+\alpha+2n$ exist above the two-body atomic (or "ionic") states ${}^{14}{\rm C}+\alpha$ in the case of $4\alpha$ system, while the two-body atomic states ${}^{4}{\rm He}+{}^{8}{\rm He}$ exist above the molecular states $\alpha+\alpha+4n$ in the case of $2\alpha$ system \cite{ito08}.
This inversion has its origin in the difference between the $\alpha-n$ interaction and $^{12}{\rm C}-n$ interaction.
This is very analogous to the electro-negativity in a molecule.
$^{14}{\rm C}$ has much larger two-neutron separation energy than $^6$He and $^8$He, so $^{12}{\rm C}$ cluster strongly attracts two valence neutrons than $\alpha$-cluster.
This effect suggests an the extended threshold rule for neutron-rich nuclei to understand the nature of the clustering.

At rather high energy region, the $4\alpha$ linear-chain configuration generates a single rotational band (cross symbols) on the bandhead state $0^+_{9}$ which has the largest overlap with the basis shown in Fig.\ref{fig:dens+} (f) which amounts to 0.72.
Although the linear-chain of 3$\alpha$ in carbon isotopes has long been investigated, there are few works for 4$\alpha$ in oxygen isotopes \cite{chev67,hori72}.
We for the first time suggest its existence and it is considerably fascinating if it is observed.

\begin{table*}[h]
 \caption{Excitation energies (MeV) and $\alpha$-decay widths (keV) of the positive-parity states in $^{18}$O. The channel radius is $a=5.2$.}
\label{tab:width+}
\begin{center}
 \begin{ruledtabular}
  \begin{tabular}{lcccccccc} 
  \multicolumn{5}{c}{\underline{This work}}
  & \multicolumn{4}{c}{\underline{Exp.}} \\ 
   Band& $J^\pi$ & $E_x$ & $\Gamma_\alpha$ & $\theta^2_\alpha$ & $J^\pi$ & $E_x$ & $\Gamma_\alpha$ & $\theta^2_\alpha$\\
   \hline
    Ground band & $0^+_1$ & $-0.16$ & - & 0.00 & $0^+_1$ & 0 & - & - \\
    & $2^+_1$ & 0.67 & - & 0.00 & $2^+_1$ & 1.98 & - & - \\
    & $4^+_1$ & 1.60 & - & 0.00 & $4^+_1$ & 3.55 & - &  \\
    ${}^{14}{\rm C}+\alpha$ & $0^+_2$ & 5.44 & - & 0.18 & $0^+_2$ & 3.63 & - & - \\
    & $2^+_3$ & 7.02 & - & 0.17 & $2^+_3$ & 5.25 & - & - \\
    & $4^+_2$ & 9.96 & 0.2 & 0.19 & $4^+_2$ & 7.12 &  &  \\
    & $6^+_1$ & 16.00 & 830 & 0.18 & $6^+_1$ & 11.69 & 12(1) \cite{avil14} & 0.23 \\
    &                                 & & & & $(6^+)$ & 11.72(5) \cite{yang19} &  & 0.56 \\
    ${}^{14}{\rm C}+\alpha$ & $0^+_4$ & 12.15 & 184 & 0.04 & $0^+$ & 9.9(1) \cite{avil14} & 3200(800) & 1.85 \\
    (higher-nodal) & $2^+_5$ & 13.41 & 135 & 0.02 & $2^+$ & 12.21(8) \cite{avil14} & 1000(250) & 0.37 \\
    &                                 & & & & $2^+$ & 12.8(3) \cite{avil14} & 4800(400) & 1.56 \\
    & $4^+_3$ & 15.27 & 308 & 0.04 & $4^+$ & 14.77(5) \cite{avil14} & 680(50) & 0.28 \\
    ${}^{12}{\rm C}+\alpha+2n$ & $0^+_3$ & 11.34 & 11.0 & 0.00 & ($0^+$) & 7.80 \cite{oert10} &  &  \\
    & $2^+_4$ & 12.36 & 13 & 0.00 & $2^+$ & 8.22 \cite{oert10} & - & - \\
    & $2^+_6$ & 13.78 & 0.0 & 0.00 & & & & \\
    & $4^+_4$ & 15.85 & 5 & 0.00 & $4^+$ & 10.30 \cite{oert10} & & \\
    &                                 & & & & $4^+$ & 10.290(4) \cite{avil14} & 19(2) & 0.09 \\
    &                                 & & & & $4^+$ & 10.28(4) \cite{yang19} &  & 0.07 \\
    & $4^+_8$ & 19.72 & 0.0 & 0.00 & & & & \\
    & $6^+_2$ & 20.12 & 4 & 0.00 & $6^+$ & 12.56 \cite{oert10} & & \\
    & $6^+_5$ & 23.01 & 2 & 0.00 & & & & \\
    & $6^+_8$ & 25.84 & 0.0 & 0.00 & & & & \\
    ${}^{12}{\rm C}+\alpha+2n$ & $0^+_6$ & 16.07 & 53 & 0.01 & & & & \\
    ($\sigma$-orbit) & $2^+_8$ & 17.56 & 23 & 0.00 & & & & \\
    & $2^+_{11}$ & 21.39 & 0.1 & 0.00 & & & & \\
    & $4^+_9$ & 19.94 & 4 & 0.00 & & & & \\
    & $4^+_{11}$ & 23.16 & 2 & 0.00 & & & & \\
    & $6^+_4$ & 22.21 & 45 & 0.00 & & & & \\
    & $6^+_7$ & 25.13 & 1 & 0.00 & & & & \\
    $4\alpha$ linear-chain & $0^+_9$ & 26.50 & 23 & 0.00 & & & & \\
    & $2^+_{17}$ & 27.46 & 4 & 0.00 & & & & \\
    & $4^+_{15}$ & 28.55 & 0.6 & 0.00 & & & & \\
    & $6^+_{10}$ & 30.19 & 0.1 & 0.00 & & & & \\
  \end{tabular}
 \end{ruledtabular}
 \end{center}
\end{table*}

\begin{figure}[h]
 \centering
 \includegraphics[width=1.0\hsize]{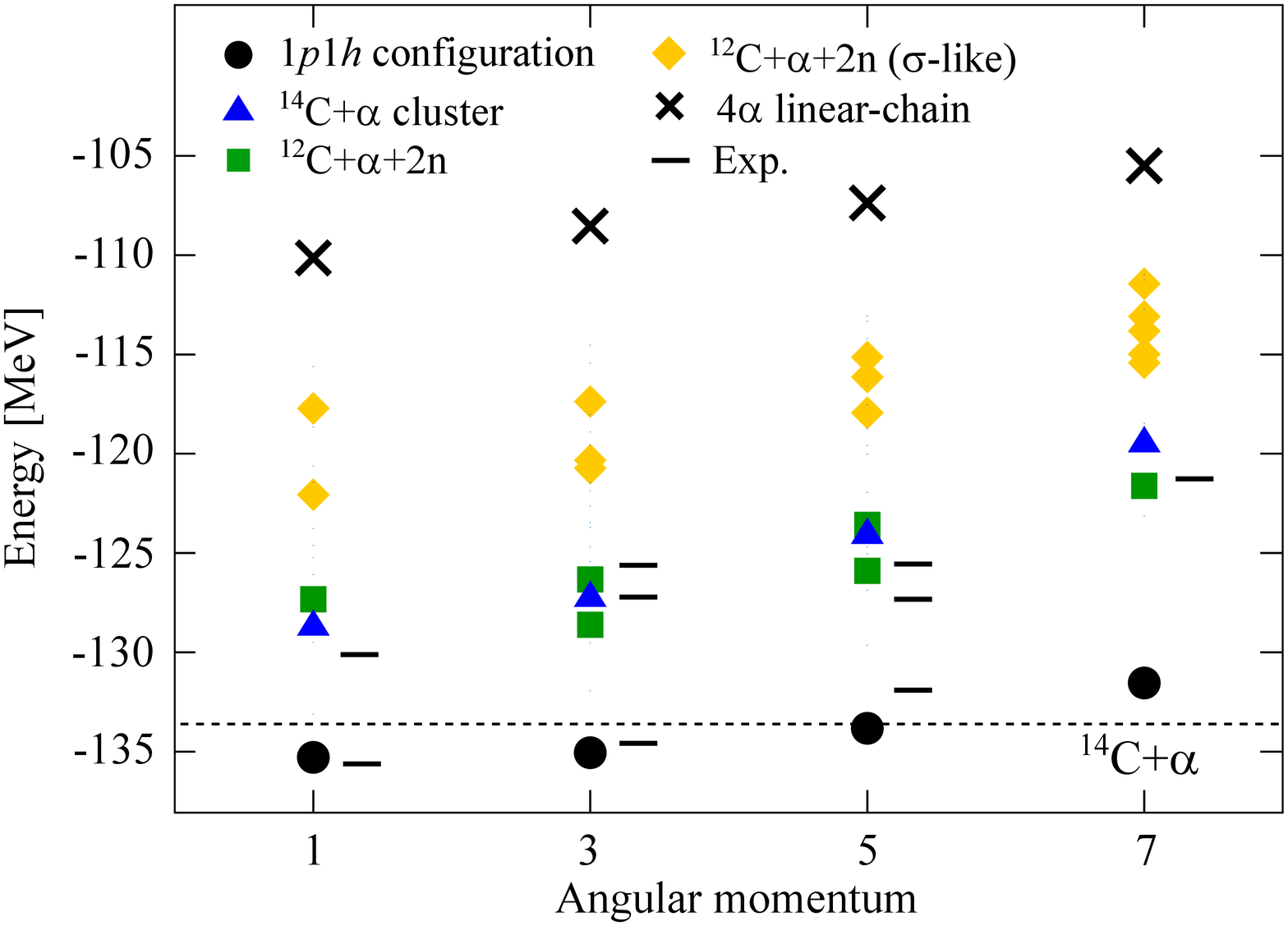}
 \caption{(color online) Negative-parity energy spectrum of $^{18}$O up to $J^\pi=7^-$. 
 Lines show the observed data \cite{avil14,yang19,endt93} and any other symbols are notated in the same manner as Figs. \ref{fig:surf-} and \ref{fig:dens-}.
 The non-cluster states are not shown.} 
 \label{fig:spec-}
\end{figure}
In the negative-parity spectrum in Fig. \ref{fig:spec-}, we discuss the five bands which correspond to the configurations seen in the energy curves.
Detailed properties and the comparison with observations are listed in Table. \ref{tab:width-}.
The ${}^{14}{\rm C}+\alpha$ configuration generates a rotational band which is built on the $1^-_4$ state located at 11.41 MeV.
The bandhead state has the largest overlap with the basis wave function shown in Fig.\ref{fig:dens-} (b) which amounts to 0.67.
The negative-parity band of ${}^{14}{\rm C}+\alpha$ is 6 MeV higher than that of positive-parity, which constitutes the inversion doublet.
The calculated bandhead of the negative-parity band is closer to the $1^-$ state in Ref. \cite{avil14} than those in Refs. \cite{gai83,gai91,curt02}.
As discussed later, the calculated $\alpha$-decay width also supports that the $1^-$ state observed by Ref. \cite{avil14} is a ${}^{14}{\rm C}+\alpha$ cluster state.

The molecular states denoted by green squares generate two rotational bands.
The $K^-=1^-$ band build on the $1^-_5$ state located at 12.49 MeV, while the $K^-=2^-$ band build on the $2^-_4$ state located at 10.43 MeV.
The bandhead state $2^-_4$ has the largest overlap with the basis wave function shown in Fig.\ref{fig:dens-} (c) which amounts to 0.87.
These member states are reasonably agreed with the excitation energies suggested by von Oertzen \cite{oert10}, although all spin-parities are tentative experimentally.
The other ${}^{12}{\rm C}+\alpha+2n$ states denoted by yellow diamonds form a $K^-=1^-$ band build on the $1^-_{10}$ state and $K^-=0^-$ band build on the $1^-_{14}$.
As the angular momentum increases, these bands are fragmented into several states due to the mixing of K quantum numbers.
The member states have large overlap with the basis wave function shown in Fig. \ref{fig:dens-} (d), which amount to, for example, 0.56 in the case of the $1^-_{10}$ state. 

Above $E_x=30$ MeV, the $4\alpha$ linear-chain band appears and the bandhead state $1^-_{16}$ has the largest overlap with the basis shown in Fig.\ref{fig:dens-} (e) which amounts to 0.89.
In the case of $^{18}$O, the linear-chain band is not fragmented and forms a single band, which is different from the negative-party linear-chain of $^{14}$C in our previous work \cite{baba16}.

\begin{table*}[h]
 \caption{Excitation energies (MeV) and $\alpha$-decay widths (keV) of the negative-parity states in $^{18}$O. The channel radius is $a=5.2$.}
\label{tab:width-}
\begin{center}
 \begin{ruledtabular}
  \begin{tabular}{lcccccccc} 
  \multicolumn{5}{c}{This work}
  & \multicolumn{4}{c}{Exp.} \\ 
   band & $J^\pi$ & $E_x$ & $\Gamma_\alpha$ & $\theta^2_\alpha$ & $J^\pi$ & $E_x$ & $\Gamma_\alpha$ & $\theta^2_\alpha$\\
   \hline
    ${}^{14}{\rm C}+\alpha$ & $1^-_4$ & 11.41 & 575 & 0.23 & $1^-$ & 9.76(2) \cite{avil14} & 630(60) & 0.46 \\
    & $3^-_7$ & 12.85 & 1367 & 0.25 & $3^-$ & 12.98(4) \cite{avil14} & 770(120) & 0.32 \\
    &                                & & & & $3^-$ & 14.0(2) \cite{avil14} & 2100(300) & 0.70 \\
    & $5^-_7$ & 16.02 & 1846 & 0.24 & $5^-$ & 12.94 \cite{avil14,yang19} & 15(2) \cite{avil14} & 0.02 \cite{avil14}, (0.50) \cite{yang19} \\
    &                                & & & & $5^-$ & 14.1 \cite{avil14,yang19} & 260(20) \cite{avil14} & 0.23 \cite{avil14}, (0.02) \cite{yang19} \\
    & $7^-_5$ & 20.63 & 2036 & 0.20 & ($7^-$) & 18.63 & &  \\
    ${}^{12}{\rm C}+\alpha+2n$ & $1^-_5$ & 12.49 & 15 & 0.00 & ($1^-$) & 10.59 \cite{oert10} & &  \\
    & $3^-_5$ & 11.20 & 5 & 0.00 & ($3^-$) & 10.92 \cite{oert10} & &  \\
    & $3^-_8$ & 13.48 & 97 & 0.02 & & & &  \\
    & $5^-_4$ & 13.92 & 2 & 0.00 & ($5^-$) & 13.83 \cite{oert10} & &  \\
    & $5^-_8$ & 16.23 & 39 & 0.01 & & & &  \\
    & $7^-_3$ & 18.19 & 1 & 0.00 & ($7^-$) & 16.98 \cite{oert10} & &  \\
    ${}^{12}{\rm C}+\alpha+2n$ & $1^-_{10}$ & 17.76 & 3 & 0.00 & & & &  \\
    (parity mixing & $3^-_{17}$ & 19.08 & 54 & 0.00 & & & &  \\
    $\quad$ $\sigma$-orbit) & $5^-_{13}$ & 21.87 & 38 & 0.00 & & & & \\
    & $7^-_7$ & 24.39 & 1 & 0.00 & & & & \\
    $4\alpha$ linear-chain & $1^-_{16}$ & 29.67 & 0.5 & 0.00 & & & & \\
    & $3^-_{25}$ & 31.27 & 0.0 & 0.00 & & & & \\
    & $5^-_{21}$ & 32.44 & 0.0 & 0.00 & & & & \\
    & $7^-_{20}$ & 34.31 & 0.0 & 0.00 & & & & \\
  \end{tabular}
 \end{ruledtabular}
 \end{center}
\end{table*}

\subsection{Decay widths}
\begin{figure}[h]
 \centering
 \includegraphics[width=1.0\hsize]{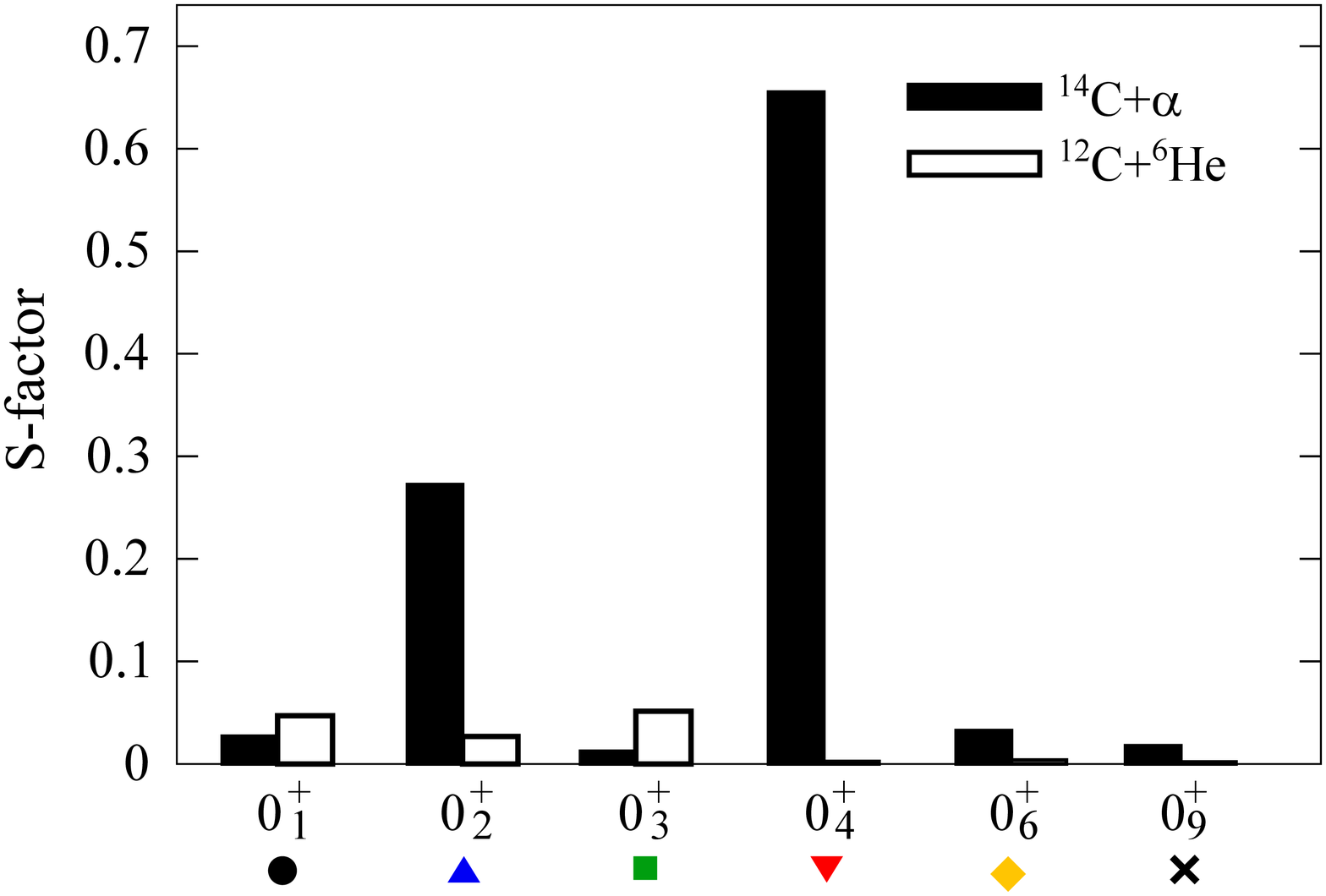}
 \caption{(color online) Spectroscopic factors with respect to the $\alpha$- and $^6$He-channels for the $J^\pi=0^+$ states.} 
 \label{fig:decay}
\end{figure}
We calculate $\alpha$-decay widths and compare with the experimental data.
Calculated widths for each band are listed in Tables. \ref{tab:width+} and \ref{tab:width-}.
The channel radius $a$ is $5.2$ fm, which is same as those used in Refs. \cite{avil14,yang19}.
The present calculation shows that the ${}^{14}{\rm C}+\alpha$ band has large dimensionless reduced $\alpha$-widths, which reasonably agrees with the observation for the $6^+_1$ state.
In addition, the higher-nodal ${}^{14}{\rm C}+\alpha$ band has large partial $\alpha$-decay widths ({\it e.g.} $\Gamma_\alpha=184$ for $0^+_4$ and $\Gamma_\alpha=308$ for $4^+_3$).
Compared with the experiments, the $0^+$ state at $E_x =9.9(1)$ MeV with the rather large $\alpha$-decay ($\Gamma_\alpha=3200(800)$) width is considered as the higher-nodal ${}^{14}{\rm C}+\alpha$ cluster state although  the excitation energy and width do not fully agree.
The underestimation of the $\alpha$-decay widths is explained as following.
As shown in Fig. \ref{fig:rwa}, the RWA of the $0^+_2$ state has the surface peak around 4.5 fm, and hence, the channel radius $a=5.2$ fm is appropriate.
On the other hand, that of the $0^+_4$ state has the surface peak around 7.0 fm.
In our calculation, therefore, the channel radius $a=5.2$ fm is not valid for the $0^+_4$ state.
Using $a=7.5$ fm, the $\alpha$-decay and dimensionless width for the $0^+_4$ state are $\Gamma_\alpha=1717$ and  $\theta^2_\alpha=0.68$, which is comparable with the observation.
In the same manner as the $0^+$, we consider the reported $2^+$ and  $4^+$ higher-lying resonances correspond to the higher-nodal ${}^{14}{\rm C}+\alpha$ band.

In the negative-parity, the reported resonances are close to the member states of the ${}^{14}{\rm C}+\alpha$ cluster band.
In particular, the calculated $1^-_4$ state at 11.41 MeV reasonably agrees with the $1^-$ state at 9.76 MeV observed by Ref. \cite{avil14}.
This implies that the bandhead of the negative-parity ${}^{14}{\rm C}+\alpha$ band is the $1^-$ state at 9.76 MeV in Ref. \cite{avil14} but not those at 4.45 and 8.04 MeV in Refs \cite{gai83,curt02,oert10}.
In order to establish the negative-parity band, we need to compare the B(E1) with the experiment \cite{gai83}.
It will be reported in a future article. 

Finally, we mention the decay patterns of the molecular states in $^{18}$O.
Figure \ref{fig:decay} shows spectroscopic factors defined by Eq. (\ref{eq:sfactor}) for the $0^+$ state.
The ${}^{14}{\rm C}+\alpha$ cluster states $0^+_2$ and $0^+_4$ show large spectroscopic factors with respect to the $\alpha$-decay (black bars).
In contrast, the ${}^{12}{\rm C}+\alpha+2n$ molecular state, $0^+_3$, shows the largest spectroscopic factor with respect to the $^6$He-decay (white bar) although it is rather smaller than $S_\alpha$ of the $0^+_2$ and $0^+_4$ states.
In addition, the $0^+_3$ state has the negligibly small $S_\alpha$.
Therefore, the $\alpha$-particle emission is dominant for the ${}^{14}{\rm C}+\alpha$ cluster states, whereas the ${}^{12}{\rm C}+\alpha+2n$ molecular states prefer the $^{6}$He emission.
This feature is consistent with the $6p4h$ configuration suggested by von Oertzen \cite{oert10}.
From the right panel of Fig. \ref{fig:rwa}, the peak of the $^6$He-decay RWA appears at the 4.5 fm.
The characteristic decay patterns can be the signature for the molecular state, if it is observed by using this channel radius.

\section{Summary}
We presented various types of clustering in ${}^{18}$O based on the AMD calculation.
It is found that the five cluster configurations appear on the energy curves for the $0^+$ state; ${}^{14}{\rm C}+\alpha$, higher-nodal ${}^{14}{\rm C}+\alpha$, ${}^{12}{\rm C}+\alpha+2n$, ${}^{12}{\rm C}+\alpha+2n$  ($\sigma$-orbit), and $4\alpha$ linear-chain states.
The calculated excitation energies of the ground band and ${}^{14}{\rm C}+\alpha$ cluster states are closer to those of the observations than previous AMD calculation due to the improvement in the effective interaction and wave functions.

We clarify the existence of the molecular state ${}^{12}{\rm C}+\alpha+2n$ suggested by von Oertzen {\it et al.}
The calculated states are reasonably close to the excited states reported by the experiment although further experimental and theoretical studies are in need.
For future observations, we also focus on the decay patterns of the obtained cluster states.
The ${}^{14}{\rm C}+\alpha$ cluster states $0^+_2$ and $0^+_4$ dominantly decay by the $\alpha$-emission.
On the other hand, the $\alpha$-decay is strongly suppressed for the ${}^{12}{\rm C}+\alpha+2n$ molecular states.
Alternatively, the $^6$He-decay gets more dominant although it shows smaller spectroscopic factor.
This characteristic decay pattern can be the signature for the molecular state.

In contrast to Be isotopes, the molecular states ${}^{12}{\rm C}+\alpha+2n$ exist above the two-body ionic states ${}^{14}{\rm C}+\alpha$ in the case of ${}^{18}$.
This inversion provides us the information of the $core-n$ interaction and affects the establishment of the extended threshold rule for neutron-rich nuclei.

The higher-nodal ${}^{14}{\rm C}+\alpha$ cluster band is the candidate of the reported resonances with rather large $\alpha$-decay widths in Refs. \cite{avil14,yang19}.
In addition, we support the $1^-$ state located at $E_x=9.76$ MeV in Ref. \cite{avil14} as the bandhead of the negative-parity ${}^{14}{\rm C}+\alpha$ cluster band which had been controversial.
In order to establish the negative-parity band, more observables such as the B(E1) are needed to be compared in future works.

\section{Acknowledgment}
M.K. acknowledges the support by the Grants-in-Aid for Scientific Research on Innovative Areas from MEXT (Grant No. 2404:24105008) and JSPS KAKENHI Grant Number JP19K03859. 
This calculation has been done on a supercomputer at Research Center for Nuclear Physics, Osaka University.
The authors acknowledge the support of the collaborative research program 2019 at Hokkaido University.

\end{document}